\documentstyle[12pt,epsfig]{article}

\catcode`\@=11
\long\def\@makefntext#1{
\protect\noindent \hbox to 3.2pt {\hskip-.9pt  
$^{{\ninerm\@thefnmark}}$\hfil}#1\hfill}		

\def\@makefnmark{\hbox to 0pt{$^{\@thefnmark}$\hss}}  
	
\def\ps@myheadings{\let\@mkboth\@gobbletwo
\def\@oddhead{\hbox{}
\rightmark\hfil\ninerm\thepage}   
\def\@oddfoot{}\def\@evenhead{\ninerm\thepage\hfil
\leftmark\hbox{}}\def\@evenfoot{}
\def\sectionmark##1{}\def\subsectionmark##1{}}

\renewcommand{\thefootnote}{\fnsymbol{footnote}}

\newcounter{sectionc}\newcounter{subsectionc}\newcounter{subsubsectionc}
\renewcommand{\section}[1] {\vspace*{0.6cm}\addtocounter{sectionc}{1} 
\setcounter{subsectionc}{0}\setcounter{subsubsectionc}{0}\noindent 
	{\normalsize\bf\thesectionc. #1}\par\vspace*{0.4cm}}
\renewcommand{\subsection}[1] {\vspace*{0.6cm}\addtocounter{subsectionc}{1} 
	\setcounter{subsubsectionc}{0}\noindent 
	{\normalsize\it\thesectionc.\thesubsectionc. #1}\par\vspace*{0.4cm}}
\renewcommand{\subsubsection}[1]
{\vspace*{0.6cm}\addtocounter{subsubsectionc}{1}
	\noindent {\normalsize\rm\thesectionc.\thesubsectionc.\thesubsubsectionc. 
	#1}\par\vspace*{0.4cm}}

\newcounter{appendixc}
\newcounter{subappendixc}[appendixc]
\newcounter{subsubappendixc}[subappendixc]

\renewcommand{\appendix}[1] {\vspace*{0.6cm}
        \refstepcounter{appendixc}
        \setcounter{figure}{0}
        \setcounter{table}{0}
        \setcounter{equation}{0}
        \renewcommand{\thefigure}{\Alph{appendixc}.\arabic{figure}}
        \renewcommand{\thetable}{\Alph{appendixc}.\arabic{table}}
        \renewcommand{\theappendixc}{\Alph{appendixc}}
        \renewcommand{\theequation}{\Alph{appendixc}.\arabic{equation}}
        \noindent{\bf Appendix \theappendixc #1}\par\vspace*{0.4cm}}

\def\abstracts#1{{
	\centering{\begin{minipage}{12.2truecm}\footnotesize\baselineskip=12pt\noindent
	\centerline{\footnotesize ABSTRACT}\vspace*{0.3cm}
	\parindent=0pt #1
	\end{minipage}}\par}} 

\newcounter{itemlistc}
\newcounter{romanlistc}
\newcounter{alphlistc}
\newcounter{arabiclistc}

\newcommand{\fcaption}[1]{
        \refstepcounter{figure}
        \setbox\@tempboxa = \hbox{\footnotesize Fig.~\thefigure. #1}
        \ifdim \wd\@tempboxa > 6in
           {\begin{center}
        \parbox{6in}{\footnotesize\baselineskip=12pt Fig.~\thefigure. #1}
            \end{center}}
        \else
             {\begin{center}
             {\footnotesize Fig.~\thefigure. #1}
              \end{center}}
        \fi}

\newcommand{\tcaption}[1]{
        \refstepcounter{table}
        \setbox\@tempboxa = \hbox{\footnotesize Table~\thetable. #1}
        \ifdim \wd\@tempboxa > 6in
           {\begin{center}
        \parbox{6in}{\footnotesize\baselineskip=12pt Table~\thetable. #1}
            \end{center}}
        \else
             {\begin{center}
             {\footnotesize Table~\thetable. #1}
              \end{center}}
        \fi}

\def\@citex[#1]#2{\if@filesw\immediate\write\@auxout
	{\string\citation{#2}}\fi
\def\@citea{}\@cite{\@for\@citeb:=#2\do
	{\@citea\def\@citea{,}\@ifundefined
	{b@\@citeb}{{\bf ?}\@warning
	{Citation `\@citeb' on page \thepage \space undefined}}
	{\csname b@\@citeb\endcsname}}}{#1}}

\newif\if@cghi
\def\cite{\@cghitrue\@ifnextchar [{\@tempswatrue
	\@citex}{\@tempswafalse\@citex[]}}
\def\citelow{\@cghifalse\@ifnextchar [{\@tempswatrue
	\@citex}{\@tempswafalse\@citex[]}}
\def\@cite#1#2{{$\null^{#1}$\if@tempswa\typeout
	{IJCGA warning: optional citation argument 
	ignored: `#2'} \fi}}

 1
 1
 1

\font\ninerm=cmr9

\def\o{\omega}
\def\g{\gamma}
\def\as{\alpha_s}
\def\asb{\bar{\alpha}_s}
\begin{document}

\centerline{\normalsize\bf k-FACTORIZATION AND NEXT-TO-LEADING ANOMALOUS
DIMENSIONS}
\baselineskip=22pt

\centerline{\footnotesize MARCELLO CIAFALONI}
\baselineskip=13pt
\centerline{\footnotesize\it Physics Department, Universit\`a di Firenze,
and INFN, Sezione di Firenze, Largo Enrico Fermi 2}
\baselineskip=12pt
\centerline{\footnotesize\it 50125 Firenze,Italy}
\centerline{\footnotesize E-mail: marcello.ciafaloni@cern.ch}
\vspace*{0.3cm}
\baselineskip=13pt
\vspace*{0.9cm}
\abstracts{I review recent results by Fadin, Lipatov and
collaborators and by our group, leading to the almost complete calculation of the
next-to-leading BFKL kernel, of its eigenvalues, and of the resummed gluon anomalous
dimension. Qualitative implications for small-$x$ structure functions include
consistent running coupling effects and a sizeable decrease of the Pomeron
intercept, so as to slow down the small-$x$ rise at low values of $Q^2$.}

\normalsize\baselineskip=15pt
\setcounter{footnote}{0}
\renewcommand{\thefootnote}{\alph{footnote}}
\section{Introduction}

The small-$x$ rise of structure functions at HERA $^1$ has been interpreted so
far by various authors $^{2-4}$ starting from two seemingly far-apart
standpoints.

The more traditional view (favoured by available fits $^{5,6}$ in the HERA
range) is that such a rise, at some value of $Q^2$, is due to large scaling
violations, i.e., to a singular anomalous dimension, starting from a flat input
distribution at some lower value $Q{^2}{_0}$.

The other view is that, for small enough values of $x = Q^2 /s$, a
qualitatively new phenomenon occurs, namely a power increase 
$F_2 \sim x^{-\omega_{\cal P}}$ of the structure functions, due to the ``hard
Pomeron" singularity.$^7$

On the other hand, the systematic use of high-energy, $k$-dependent
factorization $^{8,9}$ makes it clear $^{10}$ that the two views are rather two
regimes, coming from different rearrangements of the QCD perturbative series.
The link joining the two regimes is given by resummation formulas which provide
the $\alpha_s / \o$ dependence of the anomalous dimensions to all orders, where 
$\o \equiv N - 1 $ is the moment index, conjugate to the $\log x$ variable.

This analysis has been available for a long time $^{7,11}$ at leading $\log x$
level, and indeed shows a substantial difference between the two regimes just
mentioned.  On one hand, the gluon anomalous dimension departs from the DGLAP
value quite slowly,

\begin{equation}
\gamma_{gg} \simeq \gamma_{L} \left (
\frac{\bar{\alpha}_s}{\o} \right ) =
\frac{\asb}{\o} + 2 \zeta (3) \left (\frac{\asb}{\o} \right )^4
 + \cdots \, ,
\bar{\alpha}_s \equiv \frac{N_c \alpha_s}{\pi}
\label{I}
\end{equation}
and is roughly consistent with the so-called ``double-scaling" picture $^4$ of
HERA data.

On the other hand, the Pomeron singularity, at leading level, has the intercept

\begin{equation}
J_{\cal P} - 1 = \o_{\cal P} = 4 \bar{\alpha}_s \log \, 2
\label{II}
\end{equation}
which predicts a strong small-$x$ rise ($\o_{\cal P} = 0.4$ for $\alpha_s =
0.15$), not yet shown by the data$^4$, especially at low values of $Q^2$. 
Furthermore, if the scale of $\alpha_s$ is related to the hard scale $Q^2$, Eq.
(2) yields a sharper increase for lower $Q^2$ values, a trend not confirmed by
the data$^{12}$.

Therefore, the theoretical situation, at leading $\log x$ level, is somewhat
embarassing. Not only does it indicate that the Pomeron regime is still
further away, but it also calls for strong unitarity corrections$^{13}$ in
order to damp the small-$x$ rise at low values of $Q^2$.

Here I want to show how this picture is modified at next-to-leading level
$^{14-19}$, particularly by recent results $^{20,21}$ in the gluonic sector.
It turns out that the NL terms are sizeable and tend to slow down the small-$x$
rise, both at (resummed) anomalous dimension level and at Pomeron level,
especially at low values of $Q^2$.  As a consequence, the above regimes appear
now to be closer, and less unitarity corrections are required to achieve a
smooth matching with soft physics.

In the following, I will first use $k$-factorization to define the parton
densities and the related BFKL equation $^{7,14}$ (Section 2), and I will then
explain the features of the NL BFKL kernel (Section 3), which is the basis for
the results mentioned before.

\section{Parton densities and BFKL equation}

The unintegrated gluon density ${\cal F}^{g}_{\o} (k)$ is defined by factorization
at the exchanged (Regge) gluon of transverse momentum ${\bf k}$, and
satisfies the BFKL equation of Fig. 1.

The quark-sea density, in the DIS scheme, is defined by the $F_2$ structure
function itself and is coupled to the unintegrated gluon density by 
${\bf k}$-factorization of the quark loop, as in Fig. 2.  The above
prescriptions define the $Q_0$-scheme $^{22}$ for quarks and gluons.

The simple BFKL picture just mentioned is argued $^{16,18}$ to be valid up to NL
$\log x$ level, while at lower levels $s$-channel iteration will be
important also, leading to the effective action approach $^{23}$.

The leading kernel $K^{(L)}$ occurring in Fig. 1 is

\begin{equation}
\frac{\bar\alpha_{s}}{\o} \, K_{0} \left ( {\bf k}_{1}, {\bf  k}_{2} \right ) = 
\left . \frac{\bar\alpha_{s}}{\o} \, \frac{1}{{\bf q}^{2}} \right|_{R}, \;
{\bf q} = {\bf k}_{1} - {\bf k}_{2},
\label{III}
\end{equation}
where we have introduced the distribution

\begin{equation}
\left. \frac{1}{{\bf q}^{2}} \right|_{R} = 
\frac{1}{{\bf q}^{2}} \; \Theta({\bf q}^{2} - \lambda^{2}) -
\delta^{2} ({\bf q}) \;
\int_{\lambda^{2}}^{k^2_1} \;
\frac{d^2q}{
{\bf q}^{2}}
\label{IV}
\end{equation}

\begin{figure}
\vspace*{2.1cm}
\centerline{\psfig{figure=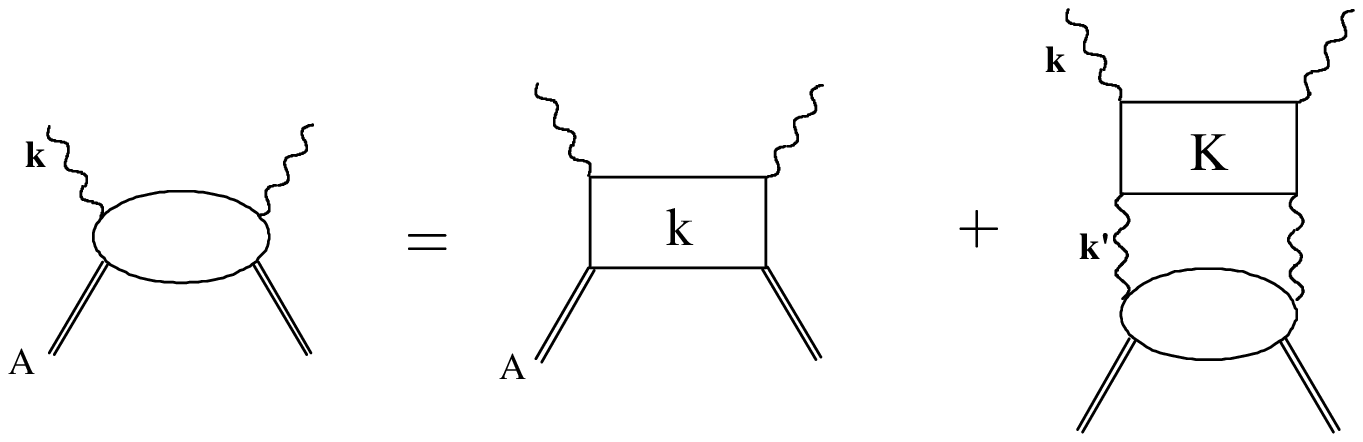}}
\vspace*{-23.7cm}
\vspace*{340pt}
\caption{\small BFKL equation for the gluon density up to 
next-to-leading level. 
Wavy lines denote (Regge) gluon exchanges, whose initial couplings to
partons $a$, $b$ are understood.}
\vspace*{-2.7cm}
\centerline{\psfig{figure=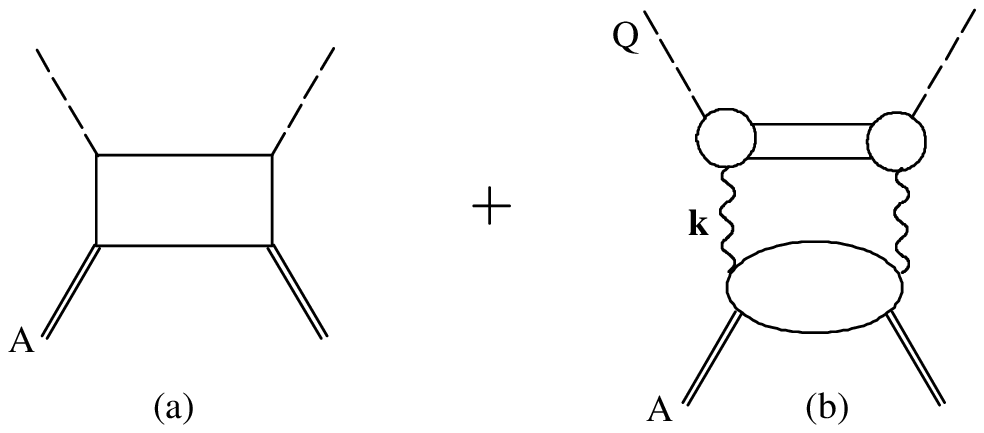}}
\vspace*{-1.3cm}
\caption{\small (a) Low energy and (b) high-energy part of the quark 
density in the 
DIS-$Q_0$ scheme. Dotted lines denote the electroweak probe and an $F_2$ 
projection is understood.}
\vspace*{-1cm}
\end{figure}

Due to scale invariance, the eigenfunctions of $K^{(L)}$ are simple powers
$\sim (k^2)^{\gamma -1}$, and the corresponding characteristic function of the
anomalous dimension has the well-known form

\begin{equation}
\chi_{0}(\gamma) = 2 \psi (1) - \psi (\gamma) - \psi (1- \gamma).
\label{V}
\end{equation}

The leading anomalous dimension 
$\gamma_{gg} = \gamma_L \left ( \frac{\bar{\alpha}_s}{\o} \right )$
is then defined by the implicit equation

\begin{equation}
1 = \frac{\bar{\alpha}_s}{\o} \; \chi_0 (\gamma_L ),
\label{VI}
\end{equation}
which provides the expansion in Eq. (1).  Furthermore, the $\o$-singularity
of $\gamma_L$ occurs around the minimum of $\chi_0$ at
$\gamma = \frac{1}{2}$, and yields by Eq. (6) the leading Pomeron
intercept in Eq. (2).

At NL level, the kernel $K = K^{(L)} + K^{(NL)}$ is explicitly dependent on the
factorization scale $\mu$, i.e., is no longer scale invariant.  I anticipate, from
the results in Section 3, that the $\mu$-dependence is logarithmic, and is dictated
by the one-loop beta function, as follows:

\begin{eqnarray}
K & = &\frac{\bar{\alpha}_s (\mu^2)}{\o} \;
\left [ \left (1 - b \, \alpha_s (\mu^2) \, \log \, \frac{k^2_1}{\mu^2} \right )
K_0 ({\bf k}_{1}, {\bf k}_{2}) +
\alpha_s (\mu^2) K_1 ({\bf k}_{1}, {\bf k}_{2}) \right ] \nonumber \\
& \simeq &
\frac{\bar{\alpha}_s (k^2_1)}{\o} \;
\left [ K_0 ({\bf k}_{1}, {\bf k}_{2}) +
\alpha_s K_1 ({\bf k}_{1}, {\bf k}_{2}) \right ],
\end{eqnarray}
where $K_0$ is the leading kernel, $12 \pi b = 11 N_c - 2 N_f$, and $K_1$ is defined
to be the scale-invariant part of the NL kernel.

The definition of $K_1$ in Eq. (7) depends on the choice of factorizing $\alpha_s$
at scale $k_1$. If another choice is made (e.g., the symmetrical one
$\alpha_s (k^2_{>})$), the form of $K_1$ changes accordingly
(e.g., $K^{>}_1 = K_1 + b\, \log  (k^2_{>} / k^2_1)$).

The form and the properties of $K_1$ are described in Section 3.  Here I just
emphasize that the NL BFKL equation, due to the lack of scale-invariance embodied
in the running coupling, needs a careful discussion $^{24,18}$ in order to relate
anomalous dimensions and the Pomeron to the characteristic function of $K_1$.

We have described $^{18}$ two small-$x$ regimes.  Anomalous dimensions can be
defined if $\o$ is not too small ($\o \to 0$ with $b\o \, 
\log ({\bf k}^2_1 / \Lambda^2
> \chi (\frac{1}{2})$).  In this regime, the larger eigenvalue
$\gamma_+  \sim \gamma_{gg} + \frac{C_F}{C_A} \, \gamma_{qg}$ of the singlet
anomalous dimension matrix is given in terms of the kernel characteristic 
function
by

\begin{equation}
1 = \frac{\bar{\alpha}_s}{\o} \, \left (\chi_0 (\gamma_+) + \alpha_s \,
\chi_1 (\gamma_+) \right )
\label{VIII}
\end{equation}
thus generalizing Eq. (6) to NL level.

By expanding Eq. (8) in $\alpha_s$, we obtain the NL resummation formula

\begin{equation}
\gamma_{gg} + \frac{C_F}{C_A} \gamma_{qg} = 
\gamma_L \left ( \frac{\bar{\alpha}_s}{w} \right ) - \alpha_s \;
\frac{\chi_1 \left (\gamma_L (\frac{\bar{\alpha}_s}{\o}) \right ) }
{\chi^\prime_0 \left (\gamma_L (\frac{\bar{\alpha}_s}{\o}) \right ) }.
\label{IX}
\end{equation}

The quark entry $\gamma_{qg}$ is provided $^{(25,21)}$ independently by 
$\bf{k}$-factorization of the quark loop in Fig. 2, namely

\begin{eqnarray}
\gamma_{qg} & = & \frac{\alpha_s N_f}{2 \pi} \, \gamma_L^2 \, h_{ab} (\gamma_L),
\nonumber \\
h_{ab} (\gamma) & = & \left ( \frac{\pi}{\sin \pi \gamma} \right)^2 \,
\frac{\cos \pi \gamma}{1-2 \gamma} \, 
\frac{1 + \frac{3}{2} \gamma (1-\gamma)}
{(1 + 2 \gamma)(3-2\gamma)}.
\label{X}
\end{eqnarray}
Therefore, the result in Eq. (9) allows the extraction of the gluon anomalous
dimension with its NL corrections, to all orders.

On the other hand, the $\o$-singularity of Eq. (8) around $\gamma_+ = \frac{1}{2}$
yields an ``$\alpha_s$-dependent Pomeron" which should be interpreted as the
singularity of the anomalous dimension expansion, separating in $\o$-space the
renormalization group regime from the Pomeron regime of very small $x$.

Since the (true) Pomeron's intercept cannot really be predicted (being dependent
$^{19}$ on the behaviour of $\alpha_s$ around ${\bf k}^2 = \Lambda^2$), the
$\o$-singularity of Eq. (8) yields a rough estimate of the effective power 
behaviour
in $x$ of $F_2$, as a function of $\alpha_s(Q^2)$.  On this basis the small-$x$
features of the structure function can be interpreted in terms of the NL
characteristic function $\chi_1 (\gamma)$.

\section{The Next-to-Leading Kernel}

The method for extracting the NL kernel was set up by Fadin, Lipatov $^{14}$ and
collaborators, and involves calculating the QCD partonic cross-section up to NL
$\log s$ level, and arranging it in a cluster expansion, which separates leading
from NL contributions (Fig. 3) which are thereby calculated $^{14-17,8,20}$.

One problem which is met in extracting the NL clusters is that one has to define an
off-shell scale for the energy in order to subtract the leading terms. Changing
such scale(s) yields an ambiguity in the definition of the NL terms.  By using a
particular prescription, Camici and I $^{21}$ have extracted the irreducible part
of the NL kernel, and we have investigated its features.

\begin{figure}
\vspace*{-1cm}
\centerline{\psfig{figure=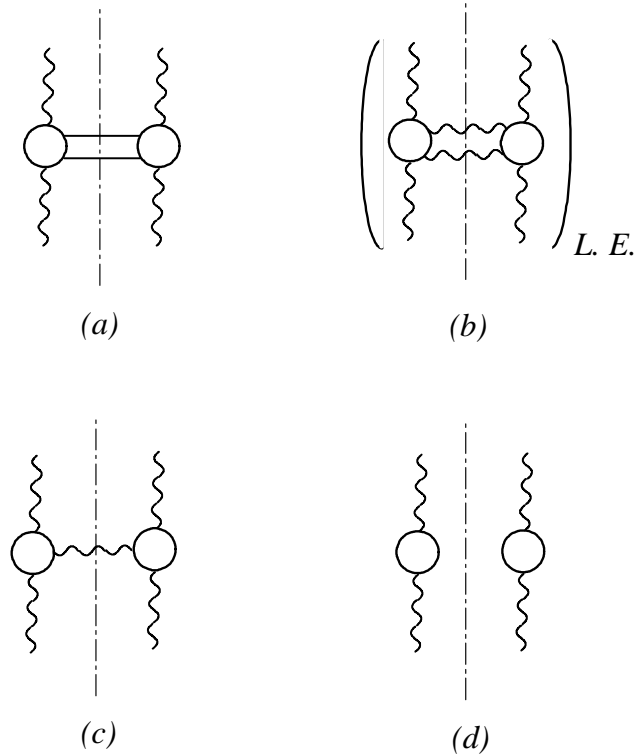,height=12cm}}
\vspace*{-1cm}
\caption{\small (a) $q\bar{q}$ and (b) 
low energy $gg$ contributions to the NL kernel,
together with (c) one-loop corrections to $1g$ state and (d) two-loop 
virtual corrections.}
\end{figure}

Referring to the original papers $^{20,21}$ for a detailed analysis, let me just
quote the results for the characteristic function $\chi_1 (\gamma)$.  The gluonic
part is 

\begin{eqnarray}
&\alpha_s \, \chi_{1}^{(g)} (\gamma) = 
\frac{\bar{\alpha}_{s}}{4} \left [ - \frac{11}{6} \,
(\chi_0^2 (\gamma) + \chi^\prime_0 (\gamma)) +
\left (\frac{67}{9} - \frac{\pi^2}{3} \right )
\chi_0 (\gamma) + \right. \nonumber \\
& +  \left. \left ( 6 \zeta (3) + \frac{\pi^2}{3\gamma (1-\gamma)} 
+ \tilde{h} (\gamma) \right ) - 
\left ( \frac{\pi}{\sin \pi \gamma} \right )^2 
\frac{\cos \pi \gamma}{3(1-2 \gamma)} 
\left ( 11 + \frac{\gamma (1-\gamma)}{(1+2\gamma) (3-2\gamma)} \right )
\right ], \hfill
\label{XI}
\end{eqnarray}
where

\begin{equation}
\tilde{h}(\gamma) \simeq \sum^{3}_{n=1}\; a_n \;
\left [ (\gamma + n)^{-1} + (1-\gamma + n)^{-1} \right ];
(a_1 = 0.72, a_2 = 0.28, a_3 = 0.16),
\label{XII}
\end{equation}
to be compared with the quark-sea part $^{18}$

\begin{equation}
\alpha_s \chi_{1}^{(q)} (\gamma) = 
\frac{N_f \alpha_s}{6 \pi}\; \left [ \frac{1}{2} 
(\chi_0^2 (\gamma) + \chi^\prime_0 (\gamma)) - 
\frac{5}{3} \chi_0 (\gamma) - \frac{1}{N_c^2} \;
h_{ab} (\gamma) \right ],
\label{XIII}
\end{equation}
where $h_{ab} (\gamma)$ is given in Eq. (10).

Consequences for the structure functions are obtained through Eqs. (8) and (9).
According to Eq. (9), the small-$\g$ behaviour of $\chi_1 (\gamma) \simeq
A_1 / \gamma^2 + A_2 / \gamma + A_3 + O(\gamma)$ is directly related to the
low-order expansion of the anomalous dimension

\begin{equation}
\gamma_{+} = \frac{\bar{\alpha}_s}{\o} + \alpha_{s}
(A_1 + A_2 \; \frac{\bar{\alpha}_s}{\o} + A_3 (\frac{\bar{\alpha}_s}{\o})^2 )
+ \cdots .
\label{XIV}
\end{equation}

Therefore, one can check that Eqs. (11) and (13) are consistent with the two-loop
results $^{26}$ in the DIS scheme

\begin{equation}
A_1 = - \frac{11 N_c}{12\pi} - \frac{N_f}{6 \pi N_c^2} ,
A_2 = - \frac{N_f}{6 \pi} \left (
\frac{5}{3} + \frac {13}{6N_c^2} \right ).
\label{XV}
\end{equation}
The three-loop entry $A_3$ is not yet available and is difficult to obtain in the
present approach also, because the scale-of-energy-dependent terms (not treated yet)
are expected to contribute.  It has been argued that coherence effects $^{27}$ in
the CCFM equation $^{27,28}$ are important for explaining $A_3$ $^{29}$.

For larger $\gamma$ values, $\chi_1 (\gamma)$ in Eq. (11) contains non-linear
effects, in particular large negative contributions, proportional to $\cos \pi
\gamma (\pi / \sin \pi \gamma)^2$, which show a $\gamma = 1$ double pole and
are
driven by the NL one-loop anomalous dimension (Fig. 4a).  When the effective
value of $\gamma \simeq \bar{\alpha}_s / \o$ increases towards the saturation
value $\gamma = \frac{1}{2}$, such terms decrease rapidly and tend to cancel in
part $^{30}$ the large scaling violations due to the leading part of
$\gamma_{gg}$ and to $\gamma_{qg}$.  

This picture is confirmed by the estimate of the $\alpha_s$-dependent Pomeron
intercept which, by Eq. (8), is given by

\begin{equation}
\o_{\cal P} (\alpha_s) = \bar{\alpha}_s (\chi_0 (\frac{1}{2}) + 
\alpha_s \chi_1 (\frac{1}{2})) = 
\bar{\alpha}_s \chi_0 (\frac{1}{2}) (1 - a \bar{\alpha}_s),
\label{XVI}
\end{equation}
where $a \simeq 3.4$ within our present knowledge (Fig. 4b).  Notice that Eq. (16)
shows a maximum $\o_{\cal P} \simeq 0.2$ at $\alpha_s = 0.15$, substantially smaller
than the leading value in Eq. (2).

\begin{figure}
\vspace*{-1.5cm}
\centerline{\hspace*{-.5cm}\psfig{figure=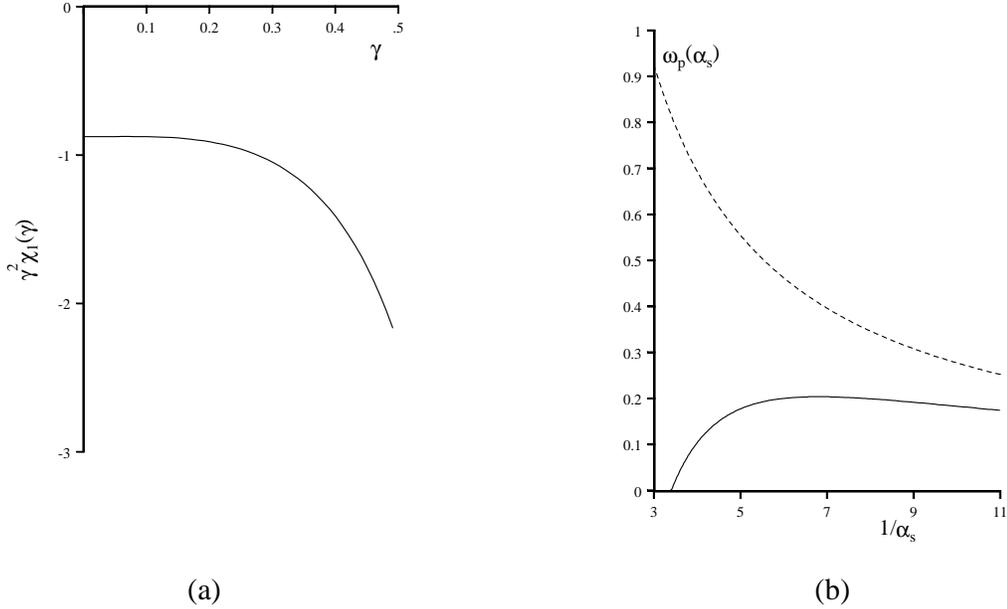,height=18cm}}
\vspace*{-9cm}
\caption{\small Plot of (a) the gluonic contribution to $\g^2\chi_1(\g)$ and
(b) the $\as$-dependent Pomeron with (full line) and without (dashed line) 
NL contributions.}
\end{figure}

Let me stress that the above indications cannot yet be taken as quantitative
estimates, due to various theoretical uncertainties (scale-of-energy-dependent
terms, higher-order collinear and coherence effects, and so on).  However, they
show a definite trend, that of slowing down the small-$x$ rise at low values of
$Q^2$.  This is satisfactory, because it reconciles somewhat the anomalous
dimension and Pomeron regimes, and helps in making a smooth transition towards soft
hadronic physics.

\section{Acknowledgements}

I wish to thank B. Kniehl for the warm atmosphere enjoyed at the Ringberg Workshop,
and J. Bartels, A. Martin, Z. Kunszt and G. Salam for a number of stimulating
discussions.  This work is supported in part by EC contract No. CHRX-CT96-0357 and
by MURST (Italy).

\section{References}

\begin{enumerate}
\item	
S. Aid et al., H1 Collaboration, {\it Nucl. Phys.} {\bf B470} (1996) 3;\\
The Zeus Collaboration, {\it Z. Phys.} {\bf C69} (1996) 607.
\item 
A.J. Askew, J. Kwiecinski, A.D. Martin and P.J. Sutton,
{\it Phys. Rev.} {\bf D47} (1993) 3775; {\bf D49} (1994) 4402;\\
C. Lopez, F. Barreiro and F.J. Yndurain, hep-ph/9605395.
\item
R.K. Ellis, F. Hautmann and B.R. Webber, {\it Phys. Lett.} {\bf B348} (1995) 582.
\item
R. Ball and S. Forte, {\it Phys. Lett.} {\bf B351} (1995) 313; {\bf B359} (1995)
362.
\item See, e.g., M. Gl\"uck, E. Reya and A. Vogt, {\it Z. Phys.} {\bf C67} (1995)
433.
\item
See, e.g., A.D. Martin, W.J. Stirling and R.G. Roberts, {\it Phys. Lett.} {\bf B387}
(1996) 419.
\item
L.N. Lipatov,{\it Sov. J. Nucl. Phys.} {\bf 23} (1976) 338; \\
E.A. Kuraev, L.N. Lipatov and V.S. Fadin, {\it Sov. Phys. JETP} {\bf 45} (1977)
199; \\
Ya. Balitskii and L.N. Lipatov, {\it Sov. J. Nucl. Phys.} {\bf 28} (1978) 822.
\item
S. Catani, M. Ciafaloni and F. Hautmann, {\it Phys. Lett.} {\bf B242} (1990)
97; {\it Nucl. Phys.} {\bf B366} (1991) 135.
\item
J.C. Collins and R.K. Ellis, {\it Nucl. Phys.} {\bf B360} (1991) 3.
\item
See, e.g., M. Ciafaloni, {\it Proc. VI-th Blois Workshop} (June 20-24, 1995).
\item
See, e.g., A. Bassetto, M. Ciafaloni and G. Marchesini, {\it Phys. Rep.}
{\bf 100} (1983) 201; 
\item
I. Bojak and M. Ernst, hep-ph/9702282.
\item
See e.g. L. V. Gribov, E. M. Levin and M. G. Ryskin, {\it Phys. Rep.}
{\bf 100} (1983) 1;
For some recent approaches, see: \\
A.H. Mueller, {\it Nucl. Phys.} {\bf B437} (1995) 107; \\
G. Salam, {\it Nucl. Phys.} {\bf B461} (1996) 512.
\item
V.S. Fadin and L.N. Lipatov, {\it Yad. Fiz.} {\bf 50} (1989) 1141.
\item
V.S. Fadin and L.N. Lipatov, {\it Nucl. Phys.} {\bf B406} (1993) 259; \\
V.S. Fadin, R. Fiore and A. Quartarolo, {\it Phys. Rev.} {\bf D50} (1994) 2265,
5893; \\
V.S. Fadin, R. Fiore and M.I. Kotsky, {\it Phys. Lett. } {\bf B389}  (1996) 737.
\item
V.S. Fadin, R. Fiore and M.I. Kotsky, {\it Phys. Lett. } {\bf B359}  (1995) 181,
{\bf B387}  (1996) 593.
\item
V.S. Fadin and L.N. Lipatov, {\it Nucl. Phys.} {\bf B477} (1996) 767.
\item
G. Camici and M. Ciafaloni, {\it Phys. Lett.} {\bf B386}  (1996) 341; 
{\it Nucl. Phys.} {\bf B496} (1997) 305.
\item
G. Camici and M. Ciafaloni, {\it Phys. Lett.} {\bf B395}  (1997) 118.
\item
V.S. Fadin, L.N. Lipatov and M.I. Kotsky, hep-ph/9704267.
\item
G. Camici and M. Ciafaloni, hep-ph/9707390.
\item
M. Ciafaloni, {\it Phys. Lett.} {\bf B356}  (1995) 74.
\item
See, e.g., L.N. Lipatov, {\it Proc. of DIS 96}, Rome, April 1996.
\item
J.C. Collins and J. Kwiecinski, {\it Nucl. Phys.} {\bf B316} (1989) 307.
\item
S. Catani and F. Hautmann, {\it Nucl. Phys.} {\bf B427} (1994) 475.
\item
G. Curci, W. Furmanski and R. Petronzio, {\it Nucl. Phys.} {\bf B175} (1980) 27; \\
E.P. Zijlstra and W.L. van Neerven, {\it Nucl. Phys.} {\bf B383} (1992) 525.
\item
M. Ciafaloni, {\it Nucl. Phys.} {\bf B296} (1988) 49, in particular Appendix C.
\item
S. Catani, F. Fiorani and G. Marchesini, {\it Nucl. Phys.} {\bf B336} (1990) 18.
\item
G. Bottazzi, G. Marchesini, G.P. Salam and M. Scorletti, hep-ph/9702418.
\item
Preliminary estimates have been made by J. Bl\"umlein and A. Vogt (private
communication).

\end{enumerate}

\end{document}